# Mid-infrared Fourier-transform spectrometer based on metamaterial lateral cladding suspended silicon waveguides

Thi Thuy Duong Dinh[1], Xavier Le Roux[1], Natnicha Koompai[1], Daniele Melati[1], Miguel Montesinos-Ballester[1], David González-Andrade[1], Pavel Cheben[2,3], Aitor V. Velasco[4], Eric Cassan[1], Delphine Marris-Morini[1], Laurent Vivien[1], Carlos Alonso-Ramos[1]

[1]Centre de Nanosciences et de Nanotechnologies, CNRS, Université Paris-Sud, Université Paris-Saclay, Palaiseau 91120, France
[2]National Research Council Canada, 1200 Montreal Road, Bldg. M50, Ottawa, Ontario K1A 0R6, Canada
[3]Center for Research in Photonics, University of Ottawa, Ottawa, Ontario K1N 6N5, Canada
[4]Instituto de Óptica Daza de Valdés, Consejo Superio de Investigaciones Científicas (CSIC), Madrid 28006, Spain
*Corresponding author: thi-thuy-duong.dinh@c2n.upsaclay.fr

**Integrated mid-infrared micro-spectrometers have a great potential for applications in environmental monitoring and space exploration. Silicon-on-insulator (SOI) is a promising platform to tackle this integration challenge, due to its unique capability for large volume and low-cost production of ultra-compact photonic circuits. However, the use of SOI in the mid-infrared is restricted by the strong absorption of the buried oxide layer for wavelengths beyond 4 µm. Here, we overcome this limitation by utilizing metamaterial-cladded suspended silicon waveguides to implement a spatial heterodyne Fourier-transform (SHFT) spectrometer operating near 5.5µm wavelength. The metamaterial-cladded geometry allows removal of the buried oxide layer, yielding measured propagation loss below 2 dB/cm between 5.3µm and 5.7µm wavelengths. The SHFT spectrometer comprises 19 Mach-Zehnder interferometers with a maximum arm length imbalance of 200 µm, achieving a measured spectral resolution of 13cm$^{-1}$ and a free-spectral range of 100 cm$^{-1}$ near 5.5µm wavelength.**

Optical spectrometers are widely used in diverse applications such as food safety, medical diagnosis, indoor air quality monitoring, astrophysics science and resource exploration [1, 2]. The mid-infrared (mid-IR, 2-20 µm wavelengths) is a particularly interesting wavelength range for spectroscopic applications as it contains the absorption fingerprints of many chemical and biological substances of interest [3, 4]. Emerging applications in planetary exploration and environmental monitoring also require the development a new generation of integrated spectrometers, providing high-sensitivity, real-time monitoring with a compact and low-cost technology [2]. In this context, multiaperture spatial heterodyne Fourier-transform (SHFT) spectrometers [5–16] provide key advantages in terms of optical throughput, resolution and robustness against fabrication imperfections, compared to dispersive devices like arrayed waveguide grating (AWGs) [17–19] and echelle gratings [20, 21]. Furthermore, the combination of the SHFT architecture and machine-learning algorithms enable advanced data analysis, yielding improved performance in terms of resolution and robustness against environmental variations such as temperature fluctuations [15, 22].

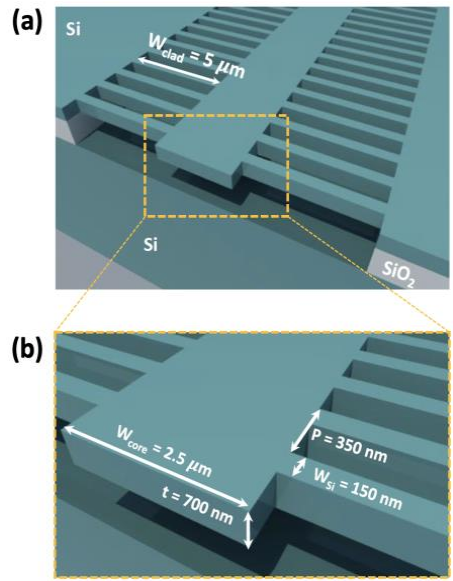

**Fig. 1.** (a) 3D Schematic view of the suspended silicon waveguide with metamaterial lateral cladding; (b) detail view showing the main geometrical parameters.

Silicon photonics have a great potential for addressing the integration challenges of compact and low-cost integrated spectrometers in the mid-IR, and particularly of SHFT spectrometers, due to its unique high-index contrast and mass-fabrication advantages [3, 4]. Mid-IR SHFT spectrometers implemented with the SiGe technology have been recently demonstrated reaching 8.5 μm wavelength with a resolution of 15 cm$^{-1}$ [11]. Yet, SHFT spectrometer demonstrations based on silicon-on-insulator (SOI) technology are limited to 3.75 μm wavelength (2 cm$^{-1}$ resolution) [9]. The main reason for this is the strong absorption of the silica cladding for wavelengths above 4 μm [4]. This limitation could be partially overcome by using the silicon-on-sapphire technology to reach 5 μm wavelength (transparency limit of sapphire). However, the only demonstration of an SHFT spectrometer in the silicon-on-sapphire technology operated near 3.3 μm wavelength, with a resolution of 10 cm$^{-1}$ [12]. Suspended silicon membrane waveguides have been identified as a promising solution to exploit the full transparency of silicon (1.1-8 μm wavelength range), as they enable the selective removal of the buried oxide layer under the waveguide core, while taking advantage of the mature SOI technology [23–30]. However, this approach has not been yet exploited for the implementation of mid-IR SHFT spectrometers in silicon.

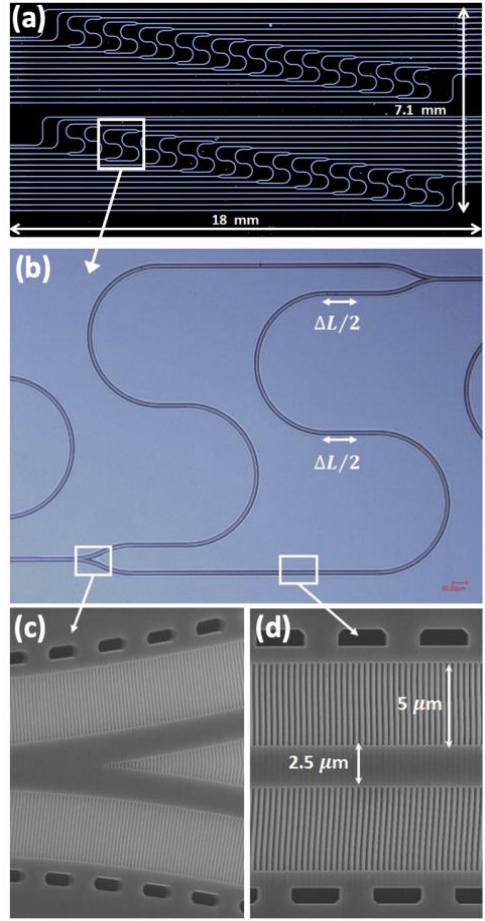

**Fig. 2.** (a) Optical images of the fabricated silicon SHFT spectrometer comprising 20 MZIs, (b) Mach-Zehnder interferometer. Scanning electron microscope images of: (c) Y-junction splitter, and (d) metamaterial-cladded suspended waveguide.

In this letter, we use suspended silicon waveguides with metamaterial-grating cladding to implement an SHFT spectrometer operating near 5.5 µm wavelength, effectively circumventing operational wavelength limitations caused by BOX absorption. The metamaterial-grating cladding provides mechanical stability and effective lateral index contrast required to confine the optical mode, while allowing fabrication with a single Si etch step [24]. The suspended waveguides, with a core thickness of 700 nm, exhibit measured propagation loss below 2 dB/cm in the 5.3-5.7 µm wavelength range. We experimentally demonstrate an SHFT spectrometer comprising 19 Mach-Zehnder interferometers (MZI) with a measured resolution of 13 cm$^{-1}$ and a free-spectral range of 120 cm$^{-1}$ near 5.5 µm wavelength.

SHFT spectrometers are generally formed by an array of MZIs, each with a different optical path length difference, thereby creating a spatial interferogram from which the input spectrum is retrieved. The wavelength resolution (dl) and the free spectral range (FSR) of the SHFT spectrometer are determined by [7]:

$$\delta\lambda = \frac{\lambda_o^2}{n_g \Delta L_{Max}}, \qquad (1)$$

$$FSR = \frac{\delta\lambda \times N}{2}, \qquad (2)$$

where $\lambda_o$ is the central wavelength, $\Delta L_{Max}$ the maximum arm length imbalance, $n_g$ is the group index of the waveguides and $N$ is the number of MZIs.

Figure 1 shows a schematic view of the proposed suspended Si waveguide with metamaterial-grating lateral cladding. The silicon thickness is 700 nm, and the buried-oxide layer thickness is 3 µm. The waveguide width of 2.5 µm and the cladding width of 5 µm minimize losses due to optical leakage to the lateral silicon slabs in the 5-6 µm wavelength range. The grating serving as lateral cladding has a period of 350 nm, operating in the subwavelength regime, i.e. with a period shorter than half of the wavelength [31]. We set a gap length of 200 nm to allow penetration of hydrofluoric (HF) acid vapor for substrate removal.

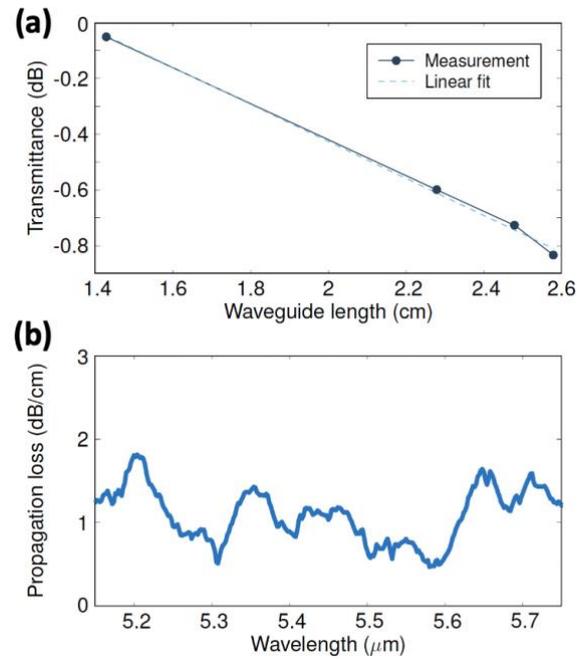

**Fig. 3.** (a) Measured transmittance for different waveguide lengths between 1.4 cm and 2.58 cm for a wavelength of 5.5 µm. Dashed line is the linear fitting, yielding a propagation loss of 0.7 dB/cm. (b) Measured propagation loss of the Si suspended waveguide as a function of the wavelength.

The SHFT spectrometer comprises 20 MZIs with optical path length linearly increasing from 20 µm to $\Delta L_{Max}$ = 200 µm (see Fig. 2(a)). The first MZI was defective and could not be used for the experiment. The complete device has a footprint of 18 mm × 7.1 mm. The interconnecting waveguides have a group index of $n_g$ = 3.8 for TE polarization, calculated with three-dimensional finite-difference time domain (FDTD) simulation, resulting in a theoretical resolution of 13 cm$^{-1}$ and an FSR of 120 cm$^{-1}$. The MZIs use Y-junctions as 1×2 splitters/combiners (measured insertion loss of 1-2 dB). The bend radius is 250 µm to ensure negligible bending loss. We fabricated the SHFT using a SOI platform with 700-nm-thick guiding Si layer and 3-µm-thick buried oxide. The patterns were defined with electron-beam lithography and dry etching. The buried oxide layer was removed using HF acid vapor [32]. The length imbalance ($\Delta L$) is implemented with two straight waveguide sections of $\Delta L$ /2 each, as indicated in Fig. 2(b). Scanning electron microscope images of the waveguide and Y-junction are shown in Figs. 2(c) and 2(d), respectively.

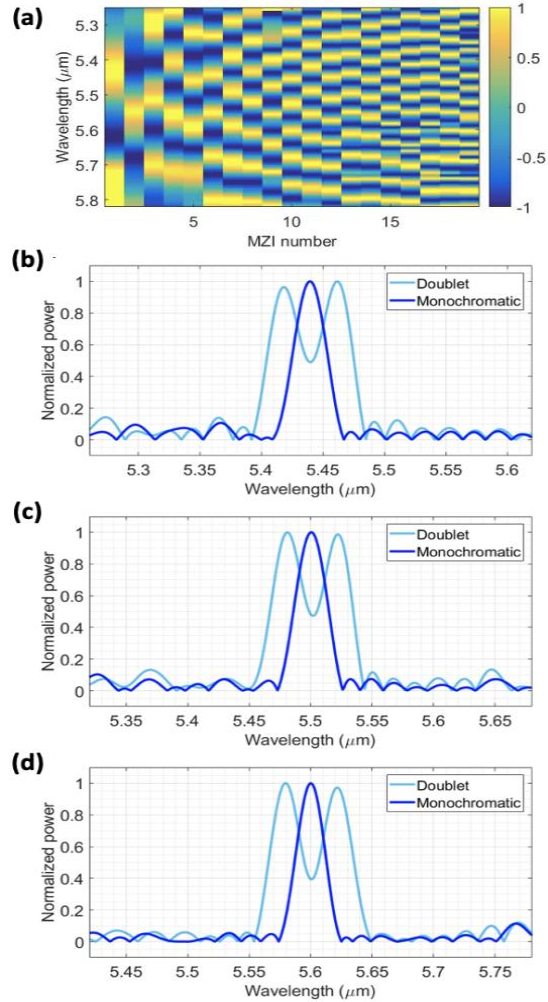

**Fig. 4.** (a) Measured calibration matrix of the SHFT spectrometer comprising the transmittance spectra of 19 MZIs. Retrieved spectrum of a monochromatic input and a doublet with peak-to-peak separation of 13 cm$^{-1}$ for wavelengths near (b) 5.45 μm, (c) 5.5 μm and (d) 5.6 μm.

To characterize the performance of the fabricated device, light from a tunable quantum cascade laser was injected and extracted from the chip using aspheric ZnSe lenses [11]. A polarization controller was used to select transverse-electric (TE) polarized light into the chip. The output signal was recorded with a mercury-cadmiun-tellurite (MCT) photodetector. We use 20- μm-wide input/output waveguides at the chip facets, with measured insertion loss of ~10 dB per facet. Propagation loss of the suspended waveguide has been determined using the cut-back method by measuring the optical transmission for 4 different waveguides with length varying between 1.4 cm and 2.58 cm. As an example, Figure 3(a) shows the transmittance as a function of the waveguide length for a wavelength of 5.5 μm. The linear fitting shows a propagation loss of 0.7 dB/cm. Figure 3(b) shows the measured propagation loss as a function of the wavelength. Our waveguides exhibit propagation losses of 1-2 dB/cm between 5.2 μm and 5.7 μm wavelength which is a substantial improvement compared to state-of-the-art metamaterial-cladded Si membrane waveguides. This improvement can be mainly attributed to optimized lithography and etching process. For comparison, previous demonstrations of metamaterial-cladded Si suspended waveguides yielded 0.8 dB/cm near 3.8 μm wavelength for 500nm Si thickness [25], 4.3 dB/cm near 6.6 μm wavelength for 1.5 μm Si thickness [28], and 3.1 dB/cm near 7.6 μm wavelength for 1.4μm Si thickness [26].

We use the pseudo-inverse transfer matrix method to calibrate the SHFT spectrometer and retrieve the input spectrum from the output interferogram, $I$, formed by the outputs of the MZIs

[7]. The pseudo-inverse retrieval method allows numerical correction of amplitude and phase errors produced by fabrication imperfections. The output interferogram $I$, can be described as $I = B \times T$, where $B$ is the input spectrum and $T$ is the calibration matrix formed by the measured transmittance of each MZI. Then, the input spectrum is retrieved by multiplying the output interferogram, $I$, by the pseudo-inverse of the calibration matrix $T$. Figure 4(a) shows the measured calibration matrix, T, comprising the transmittance spectra of the 19 MZIs near 5.5 µm wavelength. For signal retrieval we use a bandwidth of 100 cm$^{-1}$, corresponding to the theoretical FSR. In Figure 4(b-d) we plot the retrieved spectrum for monochromatic input and doublets with peak-to-peak separation of 13 cm$^{-1}$ for wavelengths near 5.44 µm (Fig. 4(b)), 5.5 µm (Fig. 4(c)) and 5.6 µm (Fig. 4(d)). The spectral resolution is ~13 cm$^{-1}$ in the three cases.

In summary, we demonstrated an integrated mid-IR SHFT spectrometer realized using suspended silicon waveguides with metamaterial-grating cladding. This strategy circumvents the absorption constrain of silica cladding for wavelengths above 4 µm, overcoming the major limitation of the SOI technology in the mid-IR. The metamaterial-cladded suspended waveguides exhibit measured propagation loss below 2 dB/cm near 5.5 µm wavelength. The SHFT spectrometer comprises an array of 19 MZIs with a maximum imbalance length of 200 µm, achieving a measured resolution of 13 cm$^{-1}$ and a bandwidth of 100 cm$^{-1}$, near 5.5 µm wavelength. This is to the best of our knowledge the longest wavelength reported for an integrated silicon SHFT spectrometer. These results open a promising route for the implementation of integrated spectrometers harnessing of the large throughput and robustness against fabrication imperfections of the SHFT architecture, exploiting the full silicon transparency range (1.1-8 µm wavelength) while taking advantage of the mature SOI technology.

**FUNDING:** French Industry Ministry (Nano2022 project under IPCEI program); Agence Nationale de la Recherche (ANR-MIRSPEC-17-CE09-0041); Spanish Ministry of Science and Innovation (RTI2018-097957-B-C33, PID2020-115353RA-I00), Community of Madrid – FEDER funds (S2018/NMT-4326).